\documentclass[preprint,showpacs,onecolumn,nofootinbib]{revtex4}
\pdfoutput=1
\usepackage[colorlinks=true,linkcolor=blue,urlcolor=blue,filecolor=black,citecolor=red,pdfstartview=FitV,pdftitle={},pdfsubject={},pdfkeywords={},pdfpagemode=None,bookmarksopen=true]{hyperref}
\usepackage{graphicx}
\usepackage{amsfonts}
\usepackage{amssymb,ulem}
\usepackage{color}%
\usepackage{dcolumn}
\usepackage{amsmath}
\usepackage{epsfig}
\usepackage{graphics}
\usepackage[active]{srcltx}
\usepackage{epstopdf}
\usepackage{shuffle}
\usepackage[utf8]{inputenc}

\newcommand{\bea}{\begin{eqnarray}}
\newcommand{\eea}{\end{eqnarray}}
\newcommand{\bean}{\begin{eqnarray*}}
\newcommand{\eean}{\end{eqnarray*}}
\newcommand{\nn}{\nonumber \\}

\def\Tr{\mathop{\rm Tr}}

\def\eref#1{(\ref{#1})}

\def\a{{\alpha}}

\def\b{{\beta}}

\def\Label#1{\label{#1}%
  \smash{\hbox to0pt{\raise1ex\hbox{\tiny[#1]}\hss}}}

\begin{document}

\baselineskip=0.6 cm
\title{Tree and $1$-loop fundamental BCJ relations from soft theorems}
\author{Fang-Stars Wei}
\email{mx120220339@stu.yzu.edu.cn}
\author{Kang Zhou}
\email{zhoukang@yzu.edu.cn}
\affiliation{Center for Gravitation and Cosmology, College of Physical Science and Technology, Yangzhou University, Yangzhou, 225009, China}

\begin{abstract}
\baselineskip=0.6 cm

We provide a new derivation of the fundamental BCJ relation among double color ordered tree amplitudes of bi-adjoint scalar theory, based on the leading soft theorem for external scalars.
Then, we generalize the fundamental BCJ relation to $1$-loop Feynman integrands. We also use the fundamental BCJ relation to understand the Adler's zero for tree amplitudes of non-linear Sigma model and Born-Infeld theories.

\end{abstract}

\maketitle


\section{Introduction}
\label{sec-intro}

As well known, soft theorems describe the universal behavior of scattering amplitudes when one or more external massless momenta are taken to near zero. Historically, soft theorems at the leading order for tree amplitudes were derived using Feynman rules \cite{Low,Weinberg}. In 2014, new soft theorems at higher orders were proposed for gravity (GR) \cite{Cachazo:2014fwa} and Yang-Mills (YM) theory \cite{Casali:2014xpa} at tree level, by applying Britto-Cachazo-Feng-Witten (BCFW) recursion relation \cite{Britto:2004ap,Britto:2005fq}. Subsequently, these new results were generalized to arbitrary space-time dimensions \cite{Schwab:2014xua,Afkhami-Jeddi:2014fia}, by using Cachazo-He-Yuan (CHY) formulas \cite{Cachazo:2013gna,Cachazo:2013hca, Cachazo:2013iea, Cachazo:2014nsa,Cachazo:2014xea}. Quite interestingly, the soft theorems can be understood as the consequence of asymptotic symmetries, and are related to memory effects \cite{Strominger:2013jfa,Strominger:2013lka,He:2014laa,Kapec:2014opa,Strominger:2014pwa,Pasterski:2015tva,Barnich1,Barnich2,Barnich3}. Furthermore, the soft theorems and asymptotic symmetries for a wide range of other theories including string theory, and the soft theorems at the loop level, have been investigated in \cite{ZviScott,HHW,FreddyEllis,Bianchi:2014gla,Campiglia:2014yka,Campiglia:2016efb,Elvang:2016qvq,Guerrieri:2017ujb,Hamada:2017atr,Mao:2017tey,
Li:2017fsb,DiVecchia:2017gfi,
Bianchi:2016viy,Chakrabarti:2017ltl,Sen:2017nim,Hamada:2018vrw}.

On the other hand, the soft theorems were exploited in the construction of tree level amplitudes, such as using another type of soft behavior called the Adler's zero to construct amplitudes of various effective theories, and the inverse soft theorem program, and so on \cite{Cheung:2014dqa,Luo:2015tat,Elvang:2018dco,Cachazo:2016njl,Rodina:2018pcb,Boucher-Veronneau:2011rwd,Nguyen:2009jk}. Using methods in above literatures, one can bootstrap amplitudes by also assuming the explicit form of soft factors or soft operators. Recently, in \cite{Zhou:2022orv}, it was shown that the tree amplitudes can be fixed by imposing only the factorization property of soft theorems and the universality of soft operators, without knowing the explicit form of such operators. The results mentioned above indicate that the tree amplitudes are completely determined by soft theorems. Thus, it is natural to expect that the relations among tree amplitudes can also be determined and understood via
soft theorems. This is the main motivation for the current note.

In this note, we use the soft theorems to derive the well known fundamental Bern-Carrasco-Johansson (BCJ) relation among color ordered amplitudes \cite{Bern:2008qj,Chiodaroli:2014xia,Johansson:2015oia,Johansson:2019dnu}. More explicitly, we use the leading order soft theorem for external scalars to get the fundamental BCJ relation among double color ordered bi-adjoint scalar (BAS) amplitudes at tree level. Furthermore, we generalize the fundamental BCJ relation to the $1$-loop level, via the forward limit method.
By expanding tree amplitudes to BAS ones, we also use the fundamental BCJ relation to understand the Adler's zero which describe the soft behavior of non-linear Sigma model (NLSM) and Born-Infeld (BI) amplitudes at tree level.

The remainder of this note is organized as follows. In section.\ref{sec-BAS}, we rapidly introduce the tree BAS amplitudes, as well as their soft behavior at the leading order. In section.\ref{sec-fund-bcj}, we derive the fundamental BCJ relation by using the leading soft theorem for external scalars. Then, we generalize the tree level fundamental BCJ relation to the $1$-loop level in section.\ref{sec-1loop}. In section.\ref{sec-Adler0}, we use the fundamental BCJ relation, as well as the expanded formula of tree NLSM and BI amplitudes, to understand the Adler's zero. Finally, we end with a summery in section.\ref{sec-conclusion}.

\section{Tree BAS amplitudes}
\label{sec-BAS}

For readers' convenience, in this section we give a brief review to the necessary background. In subsection.\ref{subsecBAS}, we introduce the tree level amplitudes of bi-adjoint scalar (BAS) theory. Some notations and technics which will be used in subsequent sections are also mentioned. In subsection.\ref{subsec-soft}, we discuss the soft behavior of tree BAS amplitudes, including the leading soft factor, as well as the statement that the leading soft behaviors of all external scalars fully determine the tree BAS amplitudes.

\subsection{Tree level BAS amplitudes}
\label{subsecBAS}

The BAS theory describes the bi-adjoint scalar field $\phi_{a\bar{a}}$ with the Lagrangian
\bea
{\cal L}_{BAS}={1\over2}\,\partial_\mu\phi^{a\bar{a}}\,\partial^{\mu}\phi^{a\bar{a}}-{\lambda\over3!}\,f^{abc}f^{\bar{a}\bar{b}\bar{c}}\,
\phi^{a\bar{a}}\phi^{b\bar{b}}\phi^{c\bar{c}}\,,
\eea
where the structure constant $f^{abc}$ and generator $T^a$ satisfy
\bea
[T^a,T^b]=if^{abc}T^c\,,
\eea
and the dual color algebra encoded by $f^{\bar{a}\bar{b}\bar{c}}$ and $T^{\bar{a}}$ is analogous.
The tree level amplitudes of this theory contain only propagators, and can be decomposed into double color ordered partial amplitudes via the standard technic.
Each double color ordered partial amplitude is simultaneously planar with respect to two color orderings, arise from expanding the full $n$-point amplitude to $\Tr(T^{a_{\sigma_1}}\cdots T^{a_{\sigma_n}})$ and $\Tr(T^{\bar{a}_{\bar{\sigma}_1}}\cdots T^{\bar{a}_{\bar{\sigma}_n}})$ respectively,
where $\sigma_i$ and $\bar{\sigma}_i$ denote permutations among all external scalars.

\begin{figure}
  \centering
  \includegraphics[width=6cm]{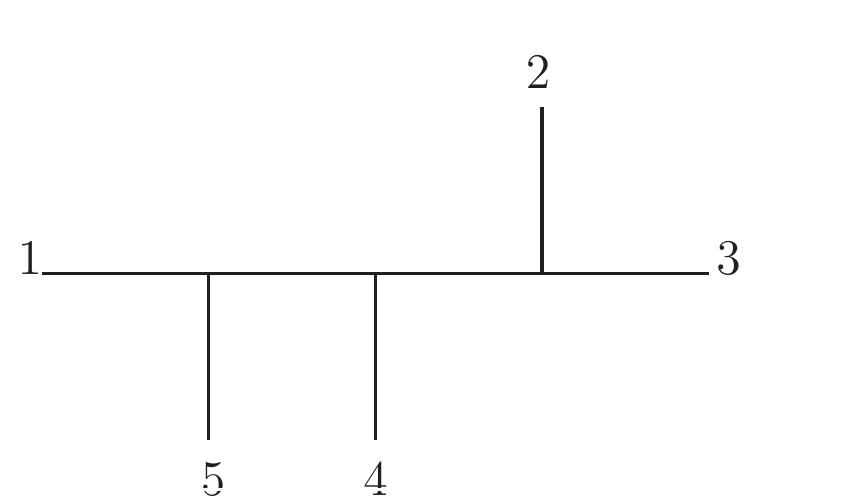}
   \includegraphics[width=6cm]{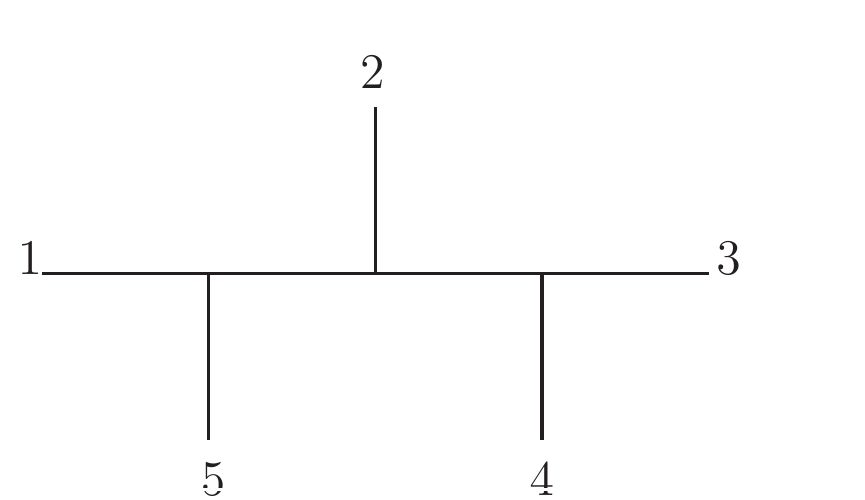}  \\
  \caption{Two $5$-point diagrams}\label{5p}
\end{figure}

\begin{figure}
  \centering
   \includegraphics[width=4cm]{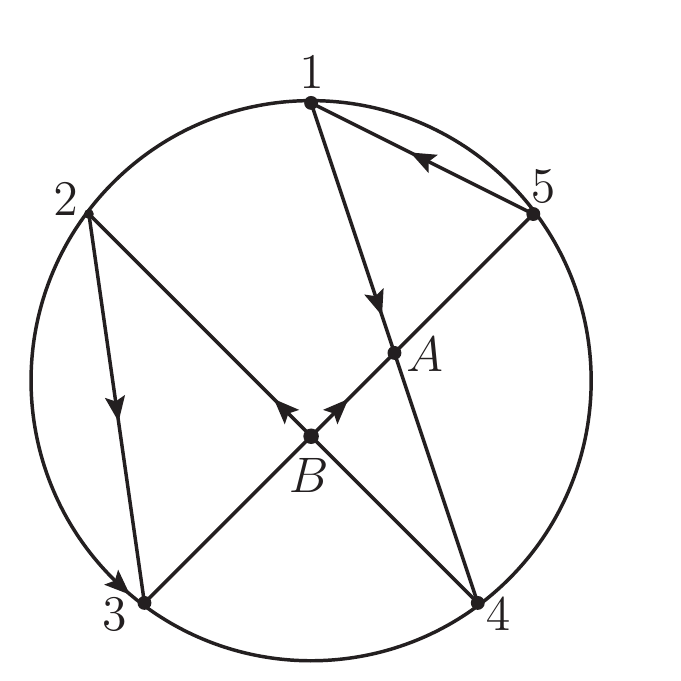}
   \includegraphics[width=4cm]{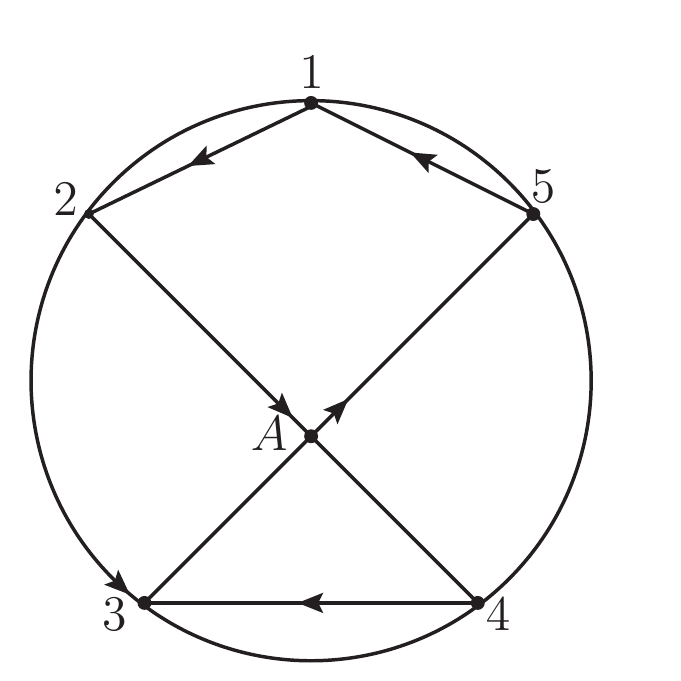} \\
  \caption{Diagram for ${\cal A}_S(1,2,3,4,5|1,4,2,3,5)$ and ${\cal A}_S(1,2,3,4,5|1,2,4,3,5)$.}\label{dis14235}
\end{figure}

To calculate double color ordered partial amplitudes, it is convenient to employ the diagrammatical method proposed by Cachazo, He and Yuan in \cite{Cachazo:2013iea}.
For a BAS amplitude whose double color-orderings are given, this method provides the corresponding Feynman diagrams as well as the overall sign directly.
Let us consider the $5$-point example ${\cal A}_S(1,2,3,4,5|1,4,2,3,5)$.
In Figure.\ref{5p}, the first diagram satisfies both two color orderings $(1,2,3,4,5)$ and $(1,4,2,3,5)$, while the second one satisfies the ordering
$(1,2,3,4,5)$ but not $(1,4,2,3,5)$. Thus, the first diagram is allowed by the double color orderings $(1,2,3,4,5|1,4,2,3,5)$, while the second one is not. It is easy to see other diagrams are also forbidden by the ordering $(1,4,2,3,5)$, thus the first diagram in Figure.\ref{5p} is the only diagram contributes to the amplitude
${\cal A}_S(1,2,3,4,5|1,4,2,3,5)$.
The Feynman diagrams contribute to a given BAS amplitude can be obtained via systematic diagrammatical rules. For the above example, one can draw a disk diagram as follows.
\begin{itemize}
\item Draw points on the boundary of the disk according to the first ordering $(1,2,3,4,5)$.
\item Draw a loop of line segments which connecting the points according to the second ordering $(1,4,2,3,5)$.
\end{itemize}
The obtained disk diagram is shown in the first diagram in Figure.\ref{dis14235}. From the diagram, one can see that two orderings share the boundaries $\{1,5\}$ and $\{2,3\}$. These co-boundaries
indicate channels ${1/s_{15}}$ and ${1/s_{23}}$, therefore the first Feynman diagram in Figure.\ref{5p}. Then the BAS amplitude ${\cal A}_S(1,2,3,4,5|1,4,2,3,5)$ can be computed as
\bea
{\cal A}_S(1,2,3,4,5|1,4,2,3,5)={1\over s_{23}}{1\over s_{51}}\,,
\eea
up to an overall sign. The Mandelstam variable $s_{i\cdots j}$ is defined as
\bea
s_{i\cdots j}\equiv K_{ij}^2\,,~~~~K_{ij}\equiv\sum_{a=i}^j\,k_a\,,~~~~\label{mandelstam}
\eea
where $k_a$ is the momentum carried by the external leg $a$.

As another example, let us consider the BAS amplitude ${\cal A}_S(1,2,3,4,5|1,2,4,3,5)$. The corresponding disk diagram is shown in the second configuration in
Figure.\ref{dis14235}, and one can see two orderings have co-boundaries $\{3,4\}$ and $\{5,1,2\}$. The co-boundary $\{3,4\}$ indicates the channel ${1/ s_{34}}$. The co-boundary $\{5,1,2\}$ indicates the channel ${1/s_{512}}$ which is equivalent to $1/s_{34}$, as well as sub-channels ${1/ s_{12}}$ and ${1/ s_{51}}$. Using the above decomposition, one can calculate ${\cal A}_S(1,2,3,4,5|1,2,4,3,5)$ as
\bea
{\cal A}_S(1,2,3,4,5|1,2,4,3,5)={1\over s_{34}}\Big({1\over s_{12}}+{1\over s_{51}}\Big)\,,
\eea
up to an overall sign.

The overall sign, determined by the color algebra, can be fixed by following rules.
\begin{itemize}
\item Each polygon with odd number of vertices contributes
a plus sign if its orientation is the same as that of the disk and a minus sign if opposite.
\item Each polygon with even number of vertices always contributes a minus sign.
\item Each intersection point contributes a minus sign.
\end{itemize}
We now apply these rules to previous examples. In the first diagram in Figure.\ref{dis14235}, the polygons are three triangles, namely $51A$, $A4B$ and $B23$, which contribute $+$, $-$, $+$ respectively, while two intersection points $A$ and $B$ contribute two $-$. In the second one in Figure.\ref{dis14235}, the polygons are $512A$ and $A43$, which contribute two $-$, while the intersection point $A$ contributes $-$. Then we arrive at the full results
\bea
{\cal A}_S(1,2,3,4,5|1,4,2,3,5)&=&-{1\over s_{23}}{1\over s_{51}}\,,\nn
{\cal A}_S(1,2,3,4,5|1,2,4,3,5)&=&-{1\over s_{34}}\Big({1\over s_{12}}+{1\over s_{51}}\Big)\,.
\eea

\begin{figure}
  \centering
   \includegraphics[width=4cm]{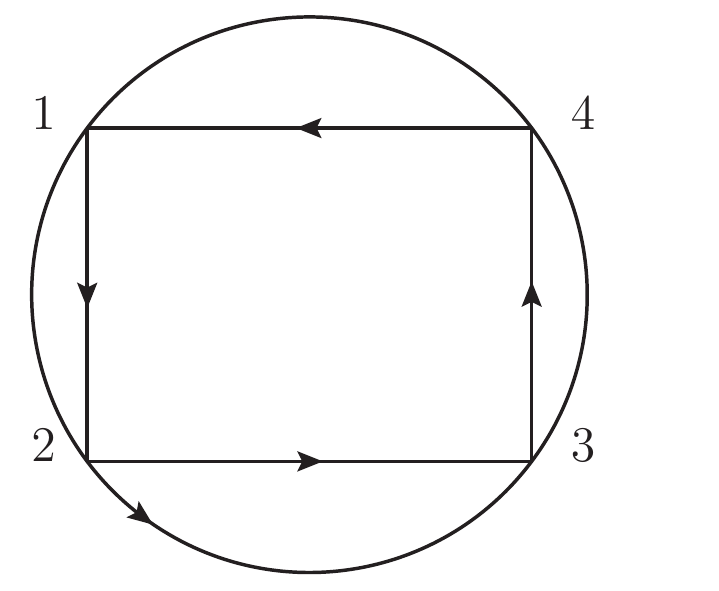} \\
  \caption{The overall sign $+$ under the new convention.}\label{newconvention}
\end{figure}

In the reminder of this paper, we adopt another convention for the overall sign. If the line segments form a convex polygon, and the orientation of the convex polygon is the same as that of the disk, then the overall sign is $+$. For instance, the disk diagram in Figure.\ref{newconvention} indicates the overall sign $+$ under the new convention, while the old convention gives $-$ according to the square formed by four line segments. Notice that the diagrammatical rules described previously still give the related sign between different disk diagrams. For example, two disk diagrams in Figure.\ref{dis14235} shows that the relative sign between ${\cal A}_S(1,2,3,4,5|1,4,2,3,5)$ and
${\cal A}_S(1,2,3,4,5|1,2,4,3,5)$ is $+$.

When considering the soft limit, the $2$-point channels play the central role. Since the partial BAS amplitude carries two color orderings, if the $2$-point channel contributes $1/s_{ab}$ to the amplitude, external legs $a$ and $b$ must be adjacent to each other in both two orderings.
Suppose the first color ordering is $(\cdots,a,b,\cdots)$, then $1/s_{ab}$ is allowed by this ordering. To denote if it is allowed by another one, we introduce the symbol $\delta_{ab}$ whose ordering of two subscripts $a$ and $b$ is determined by the first color ordering\footnote{The Kronecker symbol will not appear in this paper, thus we hope the notation $\delta_{ab}$ will not confuse the readers.}. The value of $\delta_{ab}$ is $\delta_{ab}=1$ if another color ordering is $(\cdots,a,b,\cdots)$, $\delta_{ab}=-1$ if another color ordering is $(\cdots,b,a,\cdots)$, due to the ani-symmetry of the structure constant, i.e., $f^{abc}=-f^{bac}$, and $\delta_{ab}=0$
otherwise. From the definition, it is straightforward to see $\delta_{ab}=-\delta_{ba}$. The symbol $\delta_{ab}$ will be used frequently latter.

\subsection{Leading soft behavior of tree BAS amplitudes}
\label{subsec-soft}

In this subsection, we first derive the leading order soft factor for the BAS scalar, then explain that the leading soft behaviors of all external scalars uniquely determine the tree BAS amplitudes.

The soft limit can be achieved by re-scaling
the massless momenta via a soft parameter as $k^\mu\to \tau k^\mu$, and taking the limit $\tau\to 0$. Consider the double color ordered BAS amplitude ${\cal A}_S(1,\cdots,n|\sigma)$, which carries two color orderings $(1,\cdots,n)$ and $\sigma$. We re-scale $k_i$
as $k_i\to\tau k_i$, and expand the amplitude in $\tau$. The leading order contribution manifestly aries from $2$-point channels $1/s_{1(i+1)}$
and $1/ s_{(i-1)i}$ which provide the $1/\tau$ order contributions, namely,
\bea
{\cal A}^{(0)}_S(1,\cdots,n|\sigma)&=&{1\over \tau}\Big({\delta_{i(i+1)}\over s_{i(i+1)}}+{\delta_{(i-1)i}\over s_{(i-1)i}}\Big)\,
{\cal A}_S(1,\cdots,i-1,\not{i},i+1,\cdots,n|\sigma\setminus i)\nn
&=&S^{(0)}_s(i)\,{\cal A}_S(1,\cdots,i-1,\not{i},i+1,\cdots,n|\sigma\setminus i)\,,~~~\label{for-soft-fac-s}
\eea
where $\not{i}$ stands for removing the leg $i$, $\sigma\setminus1$ means the color ordering generated from $\sigma$ by eliminating $i$. The leading soft factor $S^{(0)}_s(i)$ for the scalar $i$ is extracted as
\bea
S^{(0)}_s(i)={1\over \tau}\,\Big({\delta_{i(i+1)}\over s_{i(i+1)}}+{\delta_{(i-1)i}\over s_{(i-1)i}}\Big)\,,~~~~\label{soft-fac-s-0}
\eea
which acts on external scalars which are adjacent to $i$ in two color orderings. Notice that in our notation the parameter $\tau$ is absorbed into the soft factor $S^{(0)}_s(i)$.

If the factorization in \eref{for-soft-fac-s} is satisfied for arbitrary external leg $i\in\{1,\cdots,n\}$ when $k_i\to\tau k_i$, the tree BAS amplitude is completely determined, as can be understood as follows. The tree BAS amplitudes consist only propagators, thus are determined by correct channels. Since the soft factor \eref{soft-fac-s-0} determines the $2$-point channels, factors $S^{(0)}_s(i)$ with $i\in\{1,2,3,4\}$ manifestly fix the $4$-point BAS amplitudes.
Assume that all $(n-1)$-point BAS amplitudes are already obtained, which means all amplitudes ${\cal A}_S(1,\cdots,i-1,\not{i},i+1,\cdots,n|\sigma\setminus i)$ at the r.h.s of \eref{for-soft-fac-s} are provided. The soft theorem in \eref{for-soft-fac-s} provides all correct $2$-point poles of each $n$-point amplitude when $i$ running around $\{1,\cdots,n\}$. For higher-point channels, we observe that $1/s_{\pmb{\a}}=1/s_{\bar{\pmb{\a}}}$, due to the momentum conservation. Here $\pmb{\a}$ is a subset of $\{1,\cdots,n\}$ and $\bar{\pmb{\a}}$
is the complement, and we assume that $i\in\pmb{\a}$ where $i$ is the soft external leg. Since the pole $1/s_{\bar{\pmb{\a}}}$ is already included in the $(n-1)$-point sub-amplitude, it also contributes to the $n$-point amplitude. Thus, correct poles of $n$-point amplitude for all channels are determined, and these poles fix the $n$-point BAS amplitude uniquely. Thus, one can start from the $4$-point amplitudes which are determined by the leading soft theorem, and use the leading soft theorem to generate higher-point amplitudes recursively. Therefore, we conclude that the leading soft behaviors of all external scalars fix the tree BAS amplitudes. This conclusion is equivalent to the following statement, if a relation among tree BAS amplitudes is satisfied when taking arbitrary external leg $i\in\{1,\cdots,n\}$ to be soft, then this relation is satisfied by BAS amplitudes themselves. This inference will play the important role when deriving the fundamental BCJ relation in the next section.

\section{Fundamental BCJ relation at tree level}
\label{sec-fund-bcj}

In this section, we use the soft theorem in \eref{for-soft-fac-s} to derive the fundamental BCJ relation among double color ordered tree BAS amplitudes. In subsection.\ref{subsec-deri-fund}, we derive the fundamental BCJ relation at the leading order when one of external scalar is taken to be soft. In subsection.\ref{subsec-verify}, we show that such relation is satisfied at any order, by verifying the soft behaviors of other external scalars.

\subsection{Derivation}
\label{subsec-deri-fund}

Consider the color ordered $(n+1)$-point BAS amplitudes whose external legs are denoted as $i\in\{1,\cdots,n\}$ and $s$. The definition of $\delta_{is}$ in subsection.\ref{subsecBAS} indicates
\bea
0=\sum_{i=1}^n\,\delta_{is}\,,~~\label{rela-delta}
\eea
therefore
\bea
0&=&\sum_{i=1}^n\,(k_s\cdot k_i)\,{\delta_{is}\over s_{is}}\nn
&=&-(k_s\cdot K_{1(n-1)})\,{\delta_{ns}\over s_{ns}}+\sum_{i=1}^{n-1}\,(k_s\cdot k_i)\,{\delta_{is}\over s_{is}}\nn
&=&\sum_{i=1}^{n-1}\,(k_s\cdot k_i)\,\Big({\delta_{is}\over s_{is}}+{\delta_{sn}\over s_{sn}}\Big)\nn
&=&\sum_{i=1}^{n-1}\,\sum_{j=i}^{n-1}\,(k_s\cdot k_i)\,\Big({\delta_{js}\over s_{js}}+{\delta_{s(j+1)}\over s_{s(j+1)}}\Big)\nn
&=&\sum_{j=1}^{n-1}\,(k_s\cdot K_{1j})\,\Big({\delta_{js}\over s_{js}}+{\delta_{s(j+1)}\over s_{s(j+1)}}\Big)\,,~~~~\label{derive1}
\eea
where the combinatory momentum $K_{ab}$ is defined as $K_{ab}\equiv\sum_{i=a}^b\,k_i$, and the Mandelstam variable $s_{ij}$ is defined as $s_{ij}=2k_i\cdot k_j$. The second equality uses the momentum conservation. The third and fourth equalities are obtained by employing $\delta_{ab}=-\delta_{ba}$.
Using \eref{derive1} and the leading order soft theorem \eref{for-soft-fac-s} for the scalar $s$, we find
\bea
0=(k_s\cdot X_s)\,{\cal A}^{(0)}_{\rm S}(1,\{2,\cdots,n-1\}\shuffle s,n|\sigma)\,,~~~\label{bcj-fund-leading}
\eea
which hints the full fundamental BCJ relation
\bea
0=(k_s\cdot X_s)\,{\cal A}_{\rm S}(1,\{2,\cdots,n-1\}\shuffle s,n|\sigma)\,.~~~\label{bcj-fund}
\eea
Here the combinatory momentum $X_s$ is defined by summing over momenta carried by external scalars at the l.h.s of $s$ in the color ordering.
The symbol $\shuffle$ means summing over all
possible shuffles of two ordered sets $\pmb{\b}_1$ and $\pmb{\b}_2$, i.e., all permutations in the set $\pmb{\b}_1\cup \pmb{\b}_2$ while preserving the orderings
of $\pmb{\b}_1$ and $\pmb{\b}_2$. For instance, suppose $\pmb{\b}_1=\{1,2\}$ and $\pmb{\b}_2=\{3,4\}$, then
\bea
\pmb{\b}_1\shuffle \pmb{\b}_2=\{1,2,3,4\}+\{1,3,2,4\}+\{1,3,4,2\}+\{3,1,2,4\}+\{3,1,4,2\}+\{3,4,1,2\}\,.~~~~\label{shuffle}
\eea

The derivation in this subsection only gives \eref{bcj-fund-leading} which is the fundamental BCJ relation at the leading order, and the full one \eref{bcj-fund} should be regarded
as a conjecture at the current step. To make this work self-contained, it seems that we need to verify the relation \eref{bcj-fund} at all
orders. However, we have another more efficient choice. In subsection.\ref{subsec-soft}, we pointed out that the tree BAS amplitudes are fully
determined by leading order soft behaviors of all external scalars. The derivation of \eref{bcj-fund-leading} only uses the leading soft theorem
for the scalar $s$. Thus, we can verify the conjectured relation \eref{bcj-fund}
by checking leading soft behaviors for other external particles, as we will do in the next subsection. As will be seen, all these soft behaviors are finally reduced to the algebraic relation in \eref{rela-delta}.

The fundamental BCJ relation \eref{bcj-fund} can be generalized to color ordered YM amplitudes via the well known double copy structure.
More explicitly, the double copy structure indicates the following expansion \cite{Fu:2017uzt,Teng:2017tbo,Du:2017kpo,Du:2017gnh,Feng:2019tvb,Zhou:2019mbe}
\bea
{\cal A}_{\rm YM}(1,\cdots,n)=\sum_{\sigma'\in S_{n-2}}\,C(\sigma')\,{\cal A}_{\rm S}(1,\cdots,n|1,\sigma',n)\,.~~\label{expan-YM}
\eea
Here $S_{n-2}$ stands for permutations among $(n-2)$ legs in $\{2,\cdots,n-1\}$. The coefficients $C(\sigma')$ serve as the BCJ numerators, and
depend on permutations $\sigma'$, external momenta, as well as the polarizations of external gluons. Combining \eref{bcj-fund} and \eref{expan-YM}
together, we arrive at the relation
\bea
0=(k_s\cdot X_s)\,{\cal A}_{\rm YM}(1,\{2,\cdots,n-1\}\shuffle s,n)\,.~~~\label{bcj-fund-YM}
\eea
This is the standard BCJ relation in literatures \cite{Bern:2008qj}. One can also generate the general BCJ relations from the fundamental one, see in \cite{Ma:2011um}.

\subsection{Verification}
\label{subsec-verify}

In this subsection, we prove the fundamental BCJ relation \eref{bcj-fund} by considering the leading order soft behavior of each external scalar.
More explicitly, we will show that under the re-scaling $k_i\to\tau k_i$ with $i\in\{1,\cdots,n\}$, the leading order contribution of \eref{bcj-fund} gives the fundamental BCJ relation for BAS amplitudes with less external legs. Such reduction procedure is terminated at the simplest fundamental BCJ relation for $4$-point BAS amplitudes, whose correct soft behaviors are ensured by the definition of $\delta_{ab}$.

Under the re-scaling $k_1\to\tau k_1$, the leading order contribution of \eref{bcj-fund} is given by
\bea
0={1\over\tau}\,\Big({\delta_{12}\over s_{12}}+{\delta_{n1}\over s_{n1}}\Big)\,(k_s\cdot X_s)\,{\cal A}_{\rm S}(2,\{3,\cdots,n-1\}\shuffle s,n|\sigma\setminus1)\,,~~~\label{1-soft}
\eea
where the amplitude ${\cal A}(s,2,\cdots,n||\sigma\setminus1)$ does not contribute, since $k_s\cdot X_s$ is proportional to $\tau$ when $X_s=k_1$.
The relation \eref{1-soft} requires
\bea
0=(k_s\cdot X_s)\,{\cal A}_{\rm S}(2,\{3,\cdots,n-1\}\shuffle s,n|\sigma\setminus1)\,,~~~\label{bcj-less1}
\eea
which is nothing but the fundamental BCJ relation among $n$-point BAS amplitudes ${\cal A}_{\rm S}(2,\{3,\cdots,n-1\}\shuffle s,n|\sigma\setminus1)$.

We can use the momentum conservation to rewrite the fundamental BCJ relation \eref{bcj-fund} as
\bea
0=(k_s\cdot X^{\rm R}_s)\,{\cal A}_{\rm S}(1,\{2,\cdots,n-1\}\shuffle s,n)\,,~~~\label{bcj-fund-2}
\eea
where $X^{\rm R}_s$ is defined as the summation of momenta of legs at the r.h.s of $s$ in the color ordering.
Comparing \eref{bcj-fund-2} with \eref{bcj-fund}, we see the manifest symmetry among external legs $1$ and $n$.
This symmetry, together with the result \eref{1-soft}, indicate that for $k_n\to\tau k_n$ we have
\bea
0={1\over\tau}\,\Big({\delta_{(n-1)n}\over s_{(n-1)n}}+{\delta_{n1}\over s_{n1}}\Big)\,(k_s\cdot X_s)\,{\cal A}_{\rm S}(1,\{2,\cdots,n-2\}\shuffle s,n-1|\sigma\setminus n)\,,~~~\label{bcj-less2}
\eea
which includes the fundamental BCJ relation among $n$-point BAS amplitudes ${\cal A}_{\rm S}(1,\{2,\cdots,n-2\}\shuffle s,n-1|\sigma\setminus n)$.

For $k_i\to\tau k_i$ with $i\in\{2,\cdots,n-1\}$, the leading order contribution of \eref{bcj-fund} can be separated as
\bea
0&=&{1\over\tau}\,\Big({\delta_{(i-1)i}\over s_{(i-1)i}}+{\delta_{i(i+1)}\over s_{i(i+1)}}\Big)\,(k_s\cdot X_s)\,{\cal A}_{\rm S}(1,\{2,\cdots,i-2\}\shuffle s,i-1,\not{i},i+1,\cdots,n|\sigma\setminus i)\nn
& &+{1\over\tau}\,\Big({\delta_{(i-1)i}\over s_{(i-1)i}}+{\delta_{i(i+1)}\over s_{i(i+1)}}\Big)\,(k_s\cdot X_s)\,{\cal A}_{\rm S}(1,\cdots,i-1,\not{i},i+1,\{i+2,\cdots,n-1\}\shuffle s,n|\sigma\setminus i)\nn
& &+{1\over\tau}\,\Big({\delta_{si}\over s_{si}}+{\delta_{i(i+1)}\over s_{i(i+1)}}\Big)\,(k_s\cdot X_s)\,{\cal A}_{\rm S}(1,\cdots,i-1,s,\not{i},i+1,\cdots,n|\sigma\setminus i)\nn
& &+{1\over\tau}\,\Big({\delta_{(i-1)i}\over s_{(i-1)i}}+{\delta_{is}\over s_{is}}\Big)\,(k_s\cdot X_s)\,{\cal A}_{\rm S}(1,\cdots,i-1,\not{i},s,i+1,\cdots,n|\sigma\setminus i)\,.~~~\label{i-soft}
\eea
One can adding the last two lines at the r.h.s of \eref{i-soft} to get
\bea
& &{1\over\tau}\,\Big({\delta_{si}\over s_{si}}+{\delta_{i(i+1)}\over s_{i(i+1)}}\Big)\,(k_s\cdot X_s)\,{\cal A}_{\rm S}(1,\cdots,i-1,s,\not{i},i+1,\cdots,n|\sigma\setminus i)\nn
& &+{1\over\tau}\,\Big({\delta_{(i-1)i}\over s_{(i-1)i}}+{\delta_{is}\over s_{is}}\Big)\,(k_s\cdot X_s)\,{\cal A}_{\rm S}(1,\cdots,i-1,\not{i},s,i+1,\cdots,n|\sigma\setminus i)\nn
&=&{1\over\tau}\,\Big({\delta_{(i-1)i}\over s_{(i-1)i}}+{\delta_{i(i+1)}\over s_{i(i+1)}}\Big)\,(k_s\cdot X_s)\,{\cal A}_{\rm S}(1,\cdots,i-1,s,i+1,\cdots,n|\sigma\setminus i)\,.~~~\label{i-soft2}
\eea
Substituting \eref{i-soft2} into \eref{i-soft}, we arrive at
\bea
0={1\over\tau}\,\Big({\delta_{(i-1)i}\over s_{(i-1)i}}+{\delta_{i(i+1)}\over s_{i(i+1)}}\Big)\,(k_s\cdot X_s)\,{\cal A}_{\rm S}(1,\{\{2,\cdots,n-1\}\setminus i\}\shuffle s,n|\sigma\setminus i)\,,
\eea
which can be recognized as the fundamental BCJ relations among $n$-point BAS amplitudes ${\cal A}_{\rm S}(1,\{\{2,\cdots,n-1\}\setminus i\}\shuffle s,n|\sigma\setminus i)$.

From calculations mentioned above, we see that taking $k_i\to\tau k_i$ for each $i\in\{1,\cdots,n\}$ reduces the fundamental BCJ relation
\eref{bcj-fund} among $(n+1)$-point BAS amplitudes to the same relation among $n$-point BAS ones. This reduction can be repeated recursively.
Thus, to prove the leading order soft behavior for the conjectured relation \eref{bcj-fund}, we only need to check this relation among
$4$-point BAS amplitudes ${\cal A}(1,2\shuffle s,3|\sigma)$, namely,
\bea
0=(k_s\cdot X_s)\,{\cal A}_{\rm S}(1,2\shuffle s,3|\sigma)\,.~~~\label{bcj-4p}
\eea

The $4$-point case is quiet special, since all Mandelstam variables vanish at the $\tau^0$ order when one of external legs being soft. In other words, the leading order contribution of \eref{bcj-4p} is at the $\tau^0$ order rather than the $\tau^{-1}$ order. To see this, we use the momentum conservation to rewrite \eref{bcj-4p} as
\bea
0=(k_s\cdot k_1)\,{\cal A}_{\rm S}(1,s,2,3|\sigma)-(k_2\cdot k_1)\,{\cal A}_{\rm S}(1,2,s,3|\sigma)\,,
\eea
then re-scale $k_1$ as $k_1\to\tau k_1$, and get the leading order contribution as
\bea
0&=&(k_s\cdot k_1)\,\Big({\delta_{31}\over s_{31}}+{\delta_{1s}\over s_{1s}}\Big)\,{\cal A}_{\rm S}(s,2,3|\sigma\setminus1)-(k_1\cdot k_2)\,\Big({\delta_{31}\over s_{31}}+{\delta_{12}\over s_{12}}\Big)\,{\cal A}_{\rm S}(2,s,3|\sigma\setminus1)\,.~~~\label{bcj-4p-1soft}
\eea
To verify \eref{bcj-4p-1soft}, we observe that ${\cal A}_{\rm S}(s,2,3|\sigma\setminus1)=-{\cal A}_{\rm S}(2,s,3|\sigma\setminus1)$,
which can be understood via either the antisymmetry of the structure constant of Lie group, or the well known Kleiss-Kuijf relation \cite{Kleiss:1988ne}.
This observation turns \eref{bcj-4p-1soft} to
\bea
0&=&\Big(\delta_{12}+\delta_{13}+\delta_{1s}\Big)\,{\cal A}_{\rm S}(s,2,3|\sigma\setminus1)\,,~~~\label{bcj-4p-1soft-resul}
\eea
where we have used the momentum conservation to get the coefficient of $\delta_{13}$. The relation \eref{bcj-4p-1soft-resul} is
guaranteed by $0=\delta_{12}+\delta_{13}+\delta_{1s}$, due to the definition of $\delta_{ab}$. The correct soft behavior for $k_3\to\tau k_3$ is ensured by the symmetry between legs $1$ and $3$, as discussed around \eref{bcj-fund-2}.

Next, we use the momentum conservation to rewrite \eref{bcj-4p} as
\bea
0=(k_2\cdot k_3)\,{\cal A}_{\rm S}(1,s,2,3|\sigma)-(k_2\cdot k_1)\,{\cal A}_{\rm S}(1,2,s,3|\sigma)\,,
\eea
and consider $k_2\to\tau k_2$. The leading order contribution is
\bea
0&=&(k_2\cdot k_3)\,\Big({\delta_{s2}\over s_{s2}}+{\delta_{23}\over s_{23}}\Big)\,{\cal A}_{\rm S}(1,s,3|\sigma\setminus2)
-(k_2\cdot k_1)\,\Big({\delta_{12}\over s_{12}}+{\delta_{2s}\over s_{2s}}\Big)\,{\cal A}_{\rm S}(1,s,3|\sigma\setminus2)\nn
&=&\Big(\delta_{21}+\delta_{23}+\delta_{2s}\Big)\,{\cal A}_{\rm S}(1,s,3|\sigma\setminus2)\,,~~~\label{bcj-4p-2soft-resul}
\eea
where the momentum conservation is used to obtain the coefficient of $\delta_{2s}$. Again, the relation
\eref{bcj-4p-2soft-resul} is ensured by the definition of $\delta_{ab}$.

Thus, we conclude that the relation \eref{bcj-fund}
among double color ordered BAS amplitudes is satisfied when arbitrary external scalar is taken to be soft, therefore is correct.

\section{BCJ relation at $1$-loop level}
\label{sec-1loop}

The purpose of this section is to generalize the tree level fundamental BCJ relation to the $1$-loop level. To realize the goal, we first generalize the fundamental BCJ relation to tree BAS amplitudes with two massive external scalars in subsection.\ref{subsec-2mass}. In subsection.\ref{subsec-forward}, we review the forward limit method which generates the $1$-loop Feynman integrands from tree amplitudes. Then, in subsection.\ref{subsec-1loop}, we use the result obtained in subsection.\ref{subsec-2mass}, together with the forward limit method, to obtain the fundamental BCJ relation among Feynman integrands at $1$-loop level.

\subsection{Tree level BCJ relation with two massive external states}
\label{subsec-2mass}

In the manipulation \eref{derive1}, suppose all external momenta except $k_s$ are massive, with equal mass $k_i^2=m^2$ where $i\in\{1,\cdots,n\}$,
and all $s_{is}$ are replaced by $s_{is}-m^2=2k_i\cdot k_s$, we see that the result still holds. Furthermore, the soft factor \eref{soft-fac-s-0} are also valid under the replacement $s_{is}\to s_{is}-m^2$. This observation yields the relation \eref{bcj-fund-leading} at the leading order, thus leads us to guess the fundamental BCJ relation holds for the above massive case.
However, in the current situation, one can not conclude that the tree BAS amplitude is completely fixed by soft behaviors of external scalars. First, there is only one massless scalar $s$ can be taken to be soft thus is not sufficient to fix poles for all channels. Secondly, since we have not chose the explicit structure of interaction vertices for the new theory which includes massive scalars, in principle each internal scalar can be either massless or massive, thus it is impossible to determine corresponding poles.
Consequently, the fundamental BCJ relation does not hold for this case, as can be verified directly.

However, we can restrict ourselves to a quiet special case, the external legs $1$ and $n$ are massive with $k_1^2=k_n^2=m^2$, while all other external legs are massless. Furthermore, we assume that an massive scalar propagates through the amplitude from leg $1$ to leg $n$. Let us call this path from $1$ to $n$ the massive scalar line. One can think the full amplitude as a variety of massless scalars coupled to each other and finally coupled to the massive scalar line. For this special case, the following simple argument can convince us that the fundamental BCJ relation \eref{bcj-fund} still holds. Let us use the Mandelstam variable $s_{\pmb{\a}}$ to denote the corresponding channel, where $\pmb{\a}$ is a set of external legs satisfying $\pmb{\a}\subset\{1,\cdots,n\}\cup s$. We separate all channels into two classes, one is $1\in\pmb{\a}$, $n\in\bar{\pmb{\a}}$, and another one is $1,n\in\pmb{\a}$. For the first case, $s_{\pmb{\a}}$ corresponds to the internal line which belong to the massive line, thus the propagator takes the form $1/(s_{\pmb{\a}}-m^2)$. Since there is only one massive scalar $1$ in $\pmb{\a}$, we have
$s_{\pmb{\a}}-m^2=\sum_{i,j\in\pmb{\a}}2k_i\cdot k_j$, this form is the same as that for the massless case. For the second case, we can use $s_{\bar{\pmb{\a}}}$ to instead of $s_{\pmb{\a}}$, due to the momentum conservation. In other words, for the second case, one can always chose an expression that can not "see" the massive legs $1$ and $2$. To summarize, we find that the expression of the BAS amplitude is not affected when legs $1$ and $n$ are turned to be massive. Furthermore, in the fundamental BCJ relation \eref{bcj-fund}, it seems that there is no chance for $k_1^2$ or $k_n^2$ to play any role, even if we use the momentum conservation to rewrite any $k_i\cdot k_s$. The above argument provides the strong evidence for the validity of fundamental BCJ relation for the case $k_1^2=k_n^2=m^2$,
and we numerically verified this statement until $8$-point.

Starting from the BAS amplitudes with $k_1^2=k_n^2=m^2$, one can generate the $1$-loop Feynman integrand by sewing two massive legs together via the so called forward limit method. In this procedure, one need to allow $m^2<0$, thus it is more suitable to interpret $1$ and $n$
as off-shell legs with $k_1^2=k_2^2$. Applying this sewing manipulation, one can generalize the tree level fundamental BCJ relation to the $1$-loop level, as will be seen in subsection.\ref{subsec-1loop}.

\subsection{Forward limit method}
\label{subsec-forward}

The $1$-loop Feynman integrands can be generated from the corresponding tree amplitudes, via the so called forward limit procedure.
For instance, the $1$-loop CHY formulas can be obtained by applying this operation, as studied in \cite{He:2015yua,Cachazo:2015aol,Feng:2016nrf,Feng:2019xiq}. For readers' convenience, in this subsection we introduce the general idea and features of the forward limit.

The forward limit is reached as follows:
\begin{itemize}
\item Consider a
$(n+2)$-point tree amplitude ${\cal A}_{n+2}(k_+,k_-)$ including $n$ massless legs with momenta in $\{k_1,\cdots,k_n\}$ and two off-shell legs with $k_+^2=k_-^2\neq0$.
\item Take the limit $k_{\pm}\to \pm \ell$, and glue the two corresponding legs together. we denote this manipulation as ${\cal L}$. Performing ${\cal L}$ on the tree amplitude leads to
a special tree amplitude with $k_+=-k_-=\ell$, rather than a $1$-loop level object.
\item Sum over all allowed internal states of the internal particle with loop momentum $\ell$, such as polarization
vectors or tensors, colors, flavors, and so on\footnote{For theories include gauge or flavor groups, we only discuss the color ordered partial amplitudes in this paper, thus the summations over colors or flavors are hidden. }, we denote this manipulation as ${\cal E}$.
\end{itemize}
Roughly speaking, the obtained object, times the factor $1/\ell^2$ as
\bea
{1\over \ell^2}\,{\cal F}\,{\cal A}_{n+2}(k_+^{h_+},k_-^{h_-})={1\over \ell^2}\,\sum_{h}\,{\cal A}_{n+2}(\ell^{h},-\ell^{\bar{h}})\,,~~~~\label{single-dia}
\eea
contributes to the $n$-point $1$-loop Feynman integrand ${\bf I}_n$. Here we introduced the forward limit operator
\bea
{\cal F}\equiv{\cal E}\,{\cal L}\,,
\eea
to denote the operation of taking forward limit. In this paper, we denote the $1$-loop Feynman integrands by ${\bf I}$. From now on, we will
neglect the subscript $n$ of ${\bf I}$, since we will use other manners to denote the number of external legs.

\begin{figure}
  \centering
  \includegraphics[width=14cm]{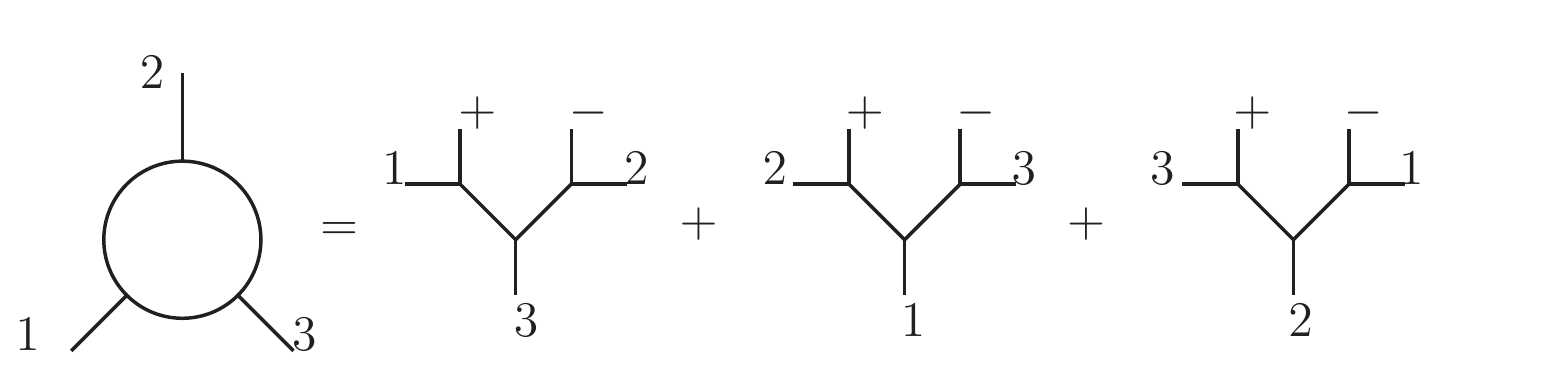} \\
  \caption{Decomposition of $1$-loop Feynman integrand.}\label{deco}
\end{figure}

For the individual Feynman diagram, the manipulation in \eref{single-dia} obviously turns the tree diagram to the $1$-loop one. However, the full
$1$-loop Feynman integrand is obtained by summing over all appropriate diagrams. Thus, let us consider what requirement should be satisfied if the resulting object of the manipulation in \eref{single-dia} can be interpreted as the correct $1$-loop Feynman integrand. It is easy to observe that after summing over all allowed tree level diagrams, each $1$-loop diagram receives contributions from tree diagrams correspond to cutting each propagator in the loop once (cutting is understood as the inverse operation of gluing legs $+$ and $-$ together, where $+$, $-$ denote external legs carry $k_+$ and $k_-$ respectively), as can be seen in Fig. \ref{deco}. Thus the statement that the operation in \eref{single-dia} generates the correct Feynman integrand holds if and only if the term for an individual $1$-loop diagram can be decomposed to terms for related tree diagrams, as shown in Fig. \ref{deco}. Such decomposition can be realized via the so called partial fraction identity \cite{He:2015yua,Baadsgaard:2015twa}:
\bea
{1\over D_1\cdots D_m}=\sum_{i=1}^m\,{1\over D_i}\Big[\prod_{j\neq i}\,{1\over D_j-D_i}\Big]\,,
\eea
which implies
\bea
{1\over \ell^2(\ell+K_1)^2(\ell+K_{12})^2\cdots (\ell+K_{1(m-1)})^2}\simeq{1\over \ell^2}\,\sum_{i=1}^m\,\Big[\prod_{j= i}^{i+m-2}\,{1\over (\ell+K_{ij})^2-\ell^2}\Big]\,.~~~~\label{loop-pro}
\eea
For each individual term at the r.h.s of the above relation, the loop momentum is shifted while result of Feynman integral is not altered.
Here $\simeq$ means the l.h.s and r.h.s are not equivalent to each other at the integrand level, but are equivalent at the integration level.
At the r.h.s of \eref{loop-pro}, we have seen the propagators with the denominates $(\ell+K_{ij})^2-\ell^2$, which are different from the standard ones $(\ell+K_{ij})^2$. This feature of propagators is the condition which should be satisfied if the manipulation in \eref{single-dia} provides the correct $1$-loop Feynman integrand. In CHY formulas, this requirement is satisfied via the $1$-loop level scattering equations. From the Feynman diagrams point of view, one can assume each propagator in the loop is massive with $m^2=\ell^2$. Thus, one can assume that the $1$-loop Feynman integrand ${\bf I}_\circ$ is obtained from the manipulation in \eref{single-dia}. To distinguish the full Feynman integrands and partial Feynman integrands obtained by decomposing the full ones via the partial fraction identity, from now on, we use ${\bf I}_\circ$ to denote the former ones, and
${\bf I}$ to denote the latter ones.

The above discussion is for the full Feynman integrands without any color ordering, and now we turn to the color ordered Feynman integrands. Since we have made sure that the full $1$-loop Feynman integrand can be generated from the full tree amplitude via the forward limit operation, let us start with the full tree amplitude. Consider a theory that external particles are in the adjoint representation of the $U(N)$ group, the full tree amplitude can be expanded
using the standard color decomposition as a sum over $(n+1)!$ terms
\bea
{\cal A}_{n+2}=\sum_{\sigma_1\in S_{n+2}/\mathbb{Z}_{n+2}}\,{\rm Tr}(T^{a_{\sigma_+}}T^{a_{\sigma_1}}\cdots T^{a_{\sigma_n}}T^{a_{\sigma_-}})
{\cal A}(\sigma_+,\sigma_1,\cdots,\sigma_n,\sigma_-)\,.~~~~\label{color-deco}
\eea
Notice that at the r.h.s it is not necessary to add the subscript $n+2$ to ${\cal A}$, since the color ordering $\sigma_+,\sigma_1,\cdots,\sigma_n,\sigma_-$ already reflects the number of external legs.
Taking the forward limit of external legs requires summing over the $U(N)$ degrees of freedom of the two internal particles. This gives rise to two kinds of terms. The first comes from permutations such that legs $+$ and $-$ are adjacent, the corresponding color factors are given as
\bea
\sum_{a_+=a_-=1}^{N^2}\,\delta_{a_+a_-}\,{\rm Tr}(T^{a_{+}}T^{a_{\sigma_1}}\cdots T^{a_{\sigma_n}}T^{a_{-}})
=N{\rm Tr}(T^{a_{\sigma_1}}\cdots T^{a_{\sigma_n}})\,,
\eea
thus contributes to the $n$-point color ordered Feynman integrand ${\bf I}_\circ(\sigma_1,\cdots,\sigma_n)$.
The second case that $+$ and $-$ are not adjacent gives rise to double-trace terms. In this paper, we only consider the single-trace terms, since the double-trace terms are determined by the single-trace ones \cite{Bern:1996je}, as can be proved by employing the tree level
Kleiss-Kuijf relation together with the forward limit operation \cite{Kleiss:1988ne}. For the single trace case, the above discussion shows that the partial integrand
obtained form taking the forward limit for ${\cal A}(+,\sigma_1,\cdots,\sigma_n,-)$ contributes to ${\bf I}_\circ(\sigma_1,\cdots,\sigma_n)$. To find the full decomposition of ${\bf I}_\circ(\sigma_1,\cdots,\sigma_n)$, we use the clear observation that several original color orderings give rise to the same trace factor after summing over $a_+$ and $a_-$, due to the cyclic symmetry of the trace factors. Collecting theses color orderings together, one finds that after taking the forward limit, the decomposition \eref{color-deco} can be organized as
\bea
{\cal F}\,{\cal A}_{n+2}=\sum_{\sigma_1\in S_{n+2}/\mathbb{Z}_{n+2}}\,{\rm Tr}(T^{a_{\sigma_1}}\cdots T^{a_{\sigma_n}})\,\sum_{j=0}^{n-1}\,{\cal F}\,
{\cal A}(+,\sigma_{1+j},\cdots,\sigma_{n+j},-)+({\rm double-trace})\,.
\eea
Consequently, the full color ordered Feynman integrand can be expanded as the following cyclic summation
\bea
{\bf I}_\circ(\sigma_1,\cdots,\sigma_n)=\sum_{j=0}^{n-1}\,
{\bf I}(+,\sigma_{1+j},\cdots,\sigma_{n+j},-)\,,~~~~\label{sum-cyclic}
\eea
where the partial color ordered integrands ${\bf I}(+,\sigma_{1+j},\cdots,\sigma_{n+j},-)$ are obtained from the color ordered tree amplitudes via the standard forward limit procedure in \eref{single-dia}, namely,
\bea
{\bf I}(+,\sigma_{1+j},\cdots,\sigma_{n+j},-)={1\over\ell^2}\,{\cal F}\,
{\cal A}(+,\sigma_{1+j},\cdots,\sigma_{n+j},-)\,.~~\label{partial}
\eea
The cyclic summation in \eref{sum-cyclic} indicates that each propagator in the loop has been cut once, thus ${\bf I}_\circ(\sigma_1,\cdots,\sigma_n)$ and $
{\bf I}(+,\sigma_{1+j},\cdots,\sigma_{n+j},-)$ are also related via the partial fraction identity.

The forward limit is well defined for the ${\cal N}=4$ SYM theory. For other theories, a quite general feature is, the obtained Feynman integrand suffer from divergence in the forward limit. Fortunately, the singular parts is found to be physically irrelevant, at least for theories under consideration in this paper. From the Feynman diagrams point of view, the singular parts generated by the forward limit correspond to tadpole diagrams, as well as babble diagrams for external legs, which do not contribute to the $S$-matrix. From the CHY point of view, the singular parts can be bypassed by employing the following observation \cite{Cachazo:2015aol}: as long as the CHY integrand is homogeneous in $\ell^\mu$,
the singular solutions contribute to the scaleless integrals which vanish under the dimensional regularization. The homogeneity in $\ell^\mu$
are satisfied by the BAS Feynman integrands under consideration in this note. Thus, in this section, we just assume that the singular parts generated by the forward limit are excluded by an appropriate way.

\subsection{Generalizing BCJ relation to $1$-loop level}
\label{subsec-1loop}

With the forward limit method introduced in subsection.\ref{subsec-forward}, now we are ready to generalize the fundamental BCJ relation to the $1$-loop level. Using \eref{sum-cyclic}
and \eref{partial}, we see that the single trace BAS Feynman integrand can be obtained through
\bea
{\bf I}_{\circ{\rm S}}(\{1,\cdots,n\}\shuffle s|\sigma)=\sum_{j=1}^{n}\,
{\bf I}_{\rm S}(+,\{j,j+1,\cdots,j-2,j-1\}\shuffle s,-|+,\sigma,-)\,,~~~~\label{sum-cyclic-bas}
\eea
with
\bea
& &{\bf I}_{\rm S}(+,\{j,j+1,\cdots,j-2,j-1\}\shuffle s,-|+,\sigma,-)\nn
&=&{1\over\ell^2}\,{\cal F}\,
{\cal A}_{\rm S}(+,\{j,j+1,\cdots,j-2,j-1\}\shuffle s,-|+,\sigma,-)\,,~~\label{partial-bas}
\eea
where $+$ and $-$ are two off-shell external legs for tree amplitudes ${\cal A}_{\rm S}(+,\{j,j+1,\cdots,j-2,j-1\}\shuffle s,-|+,\sigma,-)$, satisfying
$k_+=-k_-=\ell$.
Using the tree level fundamental BCJ relation \eref{bcj-fund}, we know that
\bea
0=(k_s\cdot X'_s)\,{\cal A}_{\rm S}(+,\{j,j+1,\cdots,j-2,j-1\}\shuffle s,-|+,\sigma,-)\,,
\eea
where $X'_s$ is defined as the summation over the loop momentum $\ell$ and momenta carried by external legs at the l.h.s of $s$ in the color ordering.
Since $k_s\cdot X'_s$ is commutable with the forward limit operator ${\cal F}$, the partial integrands ${\bf I}_{\rm S}(+,\{j,j+1,\cdots,j-2,j-1\}\shuffle s,-|+,\sigma,-)$ satisfies
\bea
0=(k_s\cdot X'_s)\,{\bf I}_{\rm S}(+,\{j,j+1,\cdots,j-2,j-1\}\shuffle s,-|+,\sigma,-)\,.~~\label{bcj-partialI}
\eea
Substituting \eref{bcj-partialI} into \eref{sum-cyclic-bas}, we arrive at
\bea
0=(k_s\cdot X'_s)\,{\bf I}_{\circ{\rm S}}(\{1,\cdots,n\}\shuffle s|\sigma)\,,~~\label{bcj-1loop}
\eea
which serves as the fundamental BCJ relation among BAS Feynman integrands at $1$-loop level.

Using the double copy structure, the relation \eref{bcj-1loop} can be generalized to YM Feynman integrands straightforwardly,
\bea
0=(k_s\cdot X'_s)\,{\bf I}_{\circ{\rm YM}}(\{1,\cdots,n\}\shuffle s)\,,~~\label{bcj-1loop-YM}
\eea
which is the same as the result found in \cite{Du:2012mt}.

\section{Understanding Adler's zero}
\label{sec-Adler0}

This section devotes to understand the Adler's zero for tree NLSM and BI amplitudes. In subsection.\ref{subsec-expan-NL-BI}, we introduce the expanded formulas of tree NLSM and BI amplitudes. Then, in subsection.\ref{subsec-Adler0}, we show that the Adler's zero can be manifested by applying the fundamental BCJ relation to the expanded formulas in subsection.\ref{subsec-expan-NL-BI}.

\subsection{Expanded NLSM and BI amplitudes}
\label{subsec-expan-NL-BI}

Among recent investigations of scattering amplitudes, one of the remarkable progresses is the expansions of amplitudes, which implies that the amplitudes of different theories can be unified, and provides a new tool for calculating amplitudes. Such expansions have been studied from various angles \cite{Fu:2017uzt,Teng:2017tbo,Du:2017kpo,Du:2017gnh,Feng:2019tvb,Zhou:2019mbe,Zhou:2022orv}. Among all of these angles, the double copy structure plays the central role. In this section, we will use the following two expansions.

The $n$-point tree NLSM amplitude ${\cal A}_{\rm N}(\sigma)$ with arbitrary color ordering $\sigma$ can be expanded to BAS amplitudes as follows \cite{Feng:2019tvb,Zhou:2019mbe},
\bea
{\cal A}_{\rm N}(\sigma)=\Big(\prod_{i=2}^{n-1}\,k_i\cdot X_i\Big)\,{\cal A}_{\rm S}(1,2\shuffle\cdots\shuffle n-1,n|\sigma)\,.~~~\label{expan-N}
\eea
The analogous expansion holds for the BI amplitudes, i.e., the $n$-point BI amplitude ${\cal A}_{\rm B}(\{1,\cdots,n\})$ can be
expanded to YM amplitudes as \cite{Feng:2019tvb,Zhou:2019mbe}
\bea
{\cal A}_{\rm B}(\{1,\cdots,n\})=\Big(\prod_{i=2}^{n-1}\,k_i\cdot X_i\Big)\,{\cal A}_{\rm Y}(1,2\shuffle\cdots\shuffle n-1,n)\,.~~~\label{expan-B}
\eea

Obviously, these two expressions of NLSM and BI amplitudes allow us to apply the fundamental BCJ relation directly.

\subsection{Adler's zero}
\label{subsec-Adler0}

The NLSM and BI amplitudes satisfy the so called Adler's zero condition, i.e., amplitudes vanish when one of external legs being soft.
Such phenomenon can be easily understood via the BCJ relation and expanded formulas in \eref{expan-N} and \eref{expan-B}.

We first show that the NLSM amplitude ${\cal A}_{\rm N}(\sigma)$ vanishes when the external leg $2$ is taken to be soft. The generalization to other leg $i$ with $i\in\{2,\cdots,n-1\}$ is straightforward. Let us consider a subset of the r.h.s of \eref{expan-N},
\bea
(k_2\cdot X_2)\,\Big(\prod_{i=3}^{n-1}\,k_{\a_i}\cdot X_{\a_i}\Big)\,{\cal A}_{\rm S}(1,2\shuffle\{\a_3,\cdots,\a_{n-1}\},n|\sigma)\,.~~~\label{2soft-1}
\eea
We define $X^{(2)}_{\a_i}$ as
\bea
X^{(2)}_{\a_i}=\lim_{k_2\to 0}\,X_{\a_i}\,,~~\label{x2}
\eea
which is a constant for ${\cal A}_{\rm S}(1,2\shuffle\{\a_3,\cdots,\a_{n-1}\},n|\sigma)$.
Then, the BCJ relation \eref{bcj-fund} imposes
\bea
0=(k_2\cdot X_2)\,\Big(\prod_{i=3}^{n-1}\,k_{\a_i}\cdot X^{(2)}_{\a_i}\Big)\,{\cal A}_{\rm S}(1,2\shuffle\{\a_3,\cdots,\a_{n-1}\},n|\sigma)\,,
\eea
since all $k_{\a_i}\cdot X^{(2)}_{\a_i}$ are constants.
Thus, the effective part of \eref{2soft-1} is
\bea
(k_2\cdot X_2)\,\Big[\Big(\prod_{i=3}^{n-1}\,k_{\a_i}\cdot X_{\a_i}\Big)-\Big(\prod_{i=3}^{n-1}\,k_{\a_i}\cdot X^{(2)}_{\a_i}\Big)\Big]\,{\cal A}_{\rm S}(1,2\shuffle\{\a_3,\cdots,\a_{n-1}\},n|\sigma)\,.
\eea
Since $X_{\a_i}-X^{(2)}_{\a_i}=0~{\rm or}~k_2$, we have
\bea
\Big(\prod_{i=3}^{n-1}\,k_{\a_i}\cdot X_{\a_i}\Big)-\Big(\prod_{i=3}^{n-1}\,k_{\a_i}\cdot X^{(2)}_{\a_i}\Big)
&=&c_1\,(k_2\cdot R_{11})+c_2\,(k_2\cdot R_{21})\,(k_2\cdot R_{22})+\cdots\,,~~~\label{power-k2}
\eea
where $R_{ij}$ denotes proper Lorentz vectors.

Now we re-scale $k_2$ as $k_2\to\tau k_2$, and expand \eref{2soft-1} by $\tau$. The leading order behavior of
${\cal A}_{\rm S}(1,2\shuffle\{\a_3,\cdots,\a_{n-1}\},n|\sigma)$ is at the $\tau^{-1}$ order. On the other hand,
the observation \eref{power-k2} shows that the effective coefficient $(k_2\cdot X_2)\,\Big[\Big(\prod_{i=3}^{n-1}\,k_{\a_i}\cdot X_{\a_i}\Big)-\Big(\prod_{i=3}^{n-1}\,k_{\a_i}\cdot X^{(2)}_{\a_i}\Big)\Big]$ is vanishing or at the $\tau^{q}$ order, where $q$ is an integer satisfying $q\geq2$. Consequently, we have
\bea
0=\lim_{\tau\to 0}\,(k_2\cdot X_2)\,\Big(\prod_{i=3}^{n-1}\,k_{\a_i}\cdot X_{\a_i}\Big)\,{\cal A}_{\rm S}(1,2\shuffle\{\a_3,\cdots,\a_{n-1}\},n|\sigma)\,,
\eea
therefore
\bea
0=\lim_{\tau\to 0}\,(k_2\cdot X_2)\,\Big(\prod_{i=3}^{n-1}\,k_{\a_i}\cdot X_{\a_i}\Big)\,{\cal A}_{\rm S}(1,2\shuffle\cdots\shuffle n-1,n|\sigma)\,.
\eea

The consideration for $k_2\to\tau k_2$ can be generalized to $k_i\to\tau k_i$ directly, with $i\in\{2,\cdots,n-1\}$. Thus all these legs
satisfy Adler's zero condition. The remaining task is to understand Adler's zero for external legs $1$ and $n$.
To deal with the case $k_1\to\tau k_1$, we consider the following subset of the r.h.s those the leg at the r.h.s of $1$ in the color ordering $(1,2\shuffle\cdots\shuffle n-1,n)$ is fixed, for instance
\bea
\Big(\prod_{i=2}^{n-1}\,k_i\cdot X_i\Big)\,{\cal A}_{\rm S}(1,2,3\shuffle\cdots\shuffle n-1,n|\sigma)\,.~~~\label{1soft-1}
\eea
Taking $k_1\to\tau k_1$ and expanding \eref{1soft-1} by $\tau$, we get the leading order contribution
\bea
(k_2\cdot k_1)\,\Big(\prod_{i=3}^{n-1}\,k_i\cdot X^{(1)}_i\Big)\,\Big({\delta_{n1}\over s_{n1}}+{\delta_{12}\over s_{12}}\Big)\,{\cal A}_{\rm S}(2,3\shuffle\cdots\shuffle n-1,n|\sigma)\,,
\eea
which vanishes due to the BCJ relation
\bea
0=\Big(\prod_{i=3}^{n-1}\,k_i\cdot X_i\Big)\,{\cal A}_{\rm S}(2,3\shuffle\cdots\shuffle n-1,n|\sigma)\,.
\eea
Here the definition of $X^{(1)}_i$ is analogous to the definition of $X^{(2)}_{\a_i}$ in \eref{x2}, i.e., $X^{(1)}_i=\lim_{k_1\to0} X_i$. Thus the combination \eref{1soft-1} satisfies the Adler's zero condition.
In the above discussion, we fixed the leg at the r.h.s of $1$ to be $2$. The analogous argument holds when
replacing $2$ by any $i\in\{2,\cdots,n-1\}$. Thus we conclude that the expanded formula \eref{expan-N} vanishes when the external leg $1$
being soft. The Adler's zero for the leg $n$ being soft can be understood via the similar manipulation, by employing the equivalent
representation \eref{bcj-fund-2} for the fundamental BCJ relation.

The Adler's zero for BI amplitudes can be understood through the paralleled procedure, by using the fundamental BCJ relation for YM amplitudes
\eref{bcj-fund-YM}.

\section{Summary}
\label{sec-conclusion}

In this note, we used the soft theorem for external scalars to derive the fundamental BCJ relation among double color ordered tree BAS amplitudes.
Then, we generalized this relation to the case that two external legs are massive. Using the fundamental BCJ relation for such special tree BAS amplitudes with massive external legs, as well as the forward limit method, we obtained the fundamental BCJ relation among BAS Feynman integrands at $1$-loop level. We also used the fundamental BCJ relation to understand the Adler's zero, which describe the soft behavior of tree NLSM and BI amplitudes.

In \cite{Ma:2011um}, one can see that the general BCJ relations can be generated from the fundamental one. It is interesting to ask whether such generalization can be realized via the soft theorems. In this note, all soft limits under consideration are those only one external leg being soft, which are called the single soft behavior. In order to obtain the general BCJ relations, it seems that we need to consider the multiple soft behaviors. Thus it is an interesting future direction.

In this work, we used the expanded formulas, as well as the fundamental BCJ relation, to understand the soft behavior of NLSM and BI amplitudes. In \cite{Zhou:2022orv}, it was shown that the expansions of single trace tree Yang-Mills-scalar and Yang-Mills amplitudes can be uniquely determined by assuming only the existence of soft theorems and the universality of soft factors, without knowing the details of soft factors. It is natural to ask, can the expansions of tree NLSM and BI amplitudes be determined by the general consideration of soft behavior? This question will be answered in our next work \cite{Zhou:20233}.

\section*{Acknowledgments}

The authors would thank Prof. Yijian Du for helpful discussions and valuable suggestions.

\end{document}